\pdfoutput=1
\documentclass[12pt,a4paper]{article}

\usepackage[T1]{fontenc}
\usepackage[utf8]{inputenc}
\usepackage{mathptmx}
\usepackage{textcomp}
\usepackage[american]{babel}

\usepackage[a4paper,margin=2.5cm]{geometry}
\usepackage{setspace}
\usepackage{graphicx}\graphicspath{{figures/}}
\usepackage{booktabs,array,multirow}
\usepackage[font=small,labelfont=bf,labelsep=period]{caption}
\usepackage{amsmath}
\usepackage{cite}
\usepackage{xcolor}
\PassOptionsToPackage{hyphens}{url}
\usepackage[hidelinks]{hyperref}
\usepackage{orcidlink}
\usepackage{enumitem}\setlist{nosep}
\hypersetup{pdftitle={Discrete quality-factor control in a side-coupled photonic crystal microcavity:
  evanescent Bloch tunnelling and the finite-cell correction},
  pdfauthor={Hasan Oguz}, pdfcreator={pdfLaTeX},
  pdfkeywords={photonic crystal, microcavity, quality factor, evanescent Bloch modes, coupled-mode theory,
  refractive-index sensing, FDTD}}
\ifdefined\pdfsuppressptexinfo\pdfsuppressptexinfo=-1\relax\fi
\ifdefined\pdftrailerid\pdftrailerid{}\fi
\newcommand{\zenodolink}{\url{https://doi.org/10.5281/zenodo.22912910}}
\newcommand{\spratrepo}{\url{https://github.com/codekyha/sprat}}
\newcommand{\supp}[1]{}

\usepackage{titlesec}
\titleformat{\section}{\normalsize\bfseries}{\thesection.}{0.6em}{}
\titleformat{\subsection}{\normalsize\bfseries\itshape}{\thesubsection.}{0.6em}{}
\titlespacing*{\section}{0pt}{1.4em}{0.6em}
\titlespacing*{\subsection}{0pt}{1.0em}{0.4em}

\newcommand{\dd}{\mathrm{d}}
\newcommand{\Nsep}{N_{\mathrm{sep}}}
\newcommand{\ncl}{n_{\mathrm{cl}}}
\newcommand{\rd}{r_{\mathrm{d}}}
\newcommand{\na}{n_{\mathrm{a}}}
\newcommand{\lamr}{\lambda_{\mathrm{r}}}
\newcommand{\Qlim}{Q_{\mathrm{lim}}}
\newcommand{\Qabs}{Q_{\mathrm{abs}}}
\newcommand{\dx}{\delta_{\mathrm{x}}}
\newcommand{\dy}{\delta_{\mathrm{y}}}

\begin{document}

% ================================================================= TITLE
\begin{center}
\parbox{0.94\textwidth}{\centering\LARGE\bfseries
Discrete quality-factor control in a side-coupled photonic crystal microcavity: evanescent Bloch
tunnelling and the finite-cell correction\par}

\vspace{1.5em}
{\large Hasan Oguz$^{1,2}$\,\orcidlink{0000-0001-7484-4415}}

\vspace{1.0em}
\parbox{0.90\textwidth}{\centering\small\itshape
$^1$\,Department of Computer Technologies, Vocational School, Istanbul Okan University,
34959 Tuzla, Istanbul, T\"{u}rkiye\\[0.25em]
$^2$\,Material Physics Simulation Laboratory, Science Faculty, Pamukkale University,
20160 Pamukkale, Denizli, T\"{u}rkiye\par}

\vspace{1.0em}
{\small E-mail: \href{mailto:hasan.oguz@okan.edu.tr}{hasan.oguz@okan.edu.tr}}
\end{center}

\vspace{0.8em}

% ================================================================= ABSTRACT
\noindent\textbf{Abstract.}
A point-defect cavity side-coupled to a line-defect waveguide in a two-dimensional photonic crystal of
silicon rods in water is studied with plane-wave expansion and finite-difference time-domain computations.
The defect-rod radius tunes the resonance continuously within the band gap, as first-order
perturbation theory predicts, whereas the quality factor changes in discrete steps set by the number of
lattice rows between cavity and guide; displacing the defect rod by up to a tenth of a period changes it by
less than 7~percent. Each added row multiplies the quality factor by a factor that follows, without an
adjustable parameter, from the decay of the evanescent Bloch channel of the bulk crystal that is
phase-matched to the guided mode; the channel at zero wavevector along the guide predicts little more than
half that factor. Two biases of a finite computational domain, leakage across the cladding and a standing
wave set up by reflections at the ends of the truncated guide, lower the per-row factor and change the quality factor by up to a factor of two if left uncorrected. Scaled to 1550~nm the design gives a
sensitivity of 634~nm per refractive index unit, which perturbation theory reproduces, and the linewidth
agrees with that of a transmission spectrum normalized to a cavity-free reference. Absorption by water caps
the quality factor near 9200 and limits the useful number of barrier rows. The guided mode lies inside the
light cone, so a slab of finite height needs a different guide.
% ABSTRACT END

\vspace{1.2em}
% ================================================================= 1
\section{Introduction}\label{sec:1}

A point defect next to a line-defect waveguide in a photonic crystal
\cite{yablonovitch1987,john1987,joannopoulos2008} forms a resonator that exchanges light with the guide by
tunneling through the lattice rows between them. Channel-drop filters \cite{fan1998,fan1999,manolatou1999}
and the trapping and emission of photons by a single defect in a photonic crystal slab \cite{noda2000} rely
on this coupling. The quality factor $Q$ of a point defect grows exponentially with the number of lattice
rows that surround it \cite{villeneuve1996}, and nanocavities engineered in photonic crystal slabs reach
quality factors far above those of simple point defects \cite{akahane2003,notomi2010}.

Inside a band gap the crystal carries only evanescent Bloch modes, and these describe its barriers,
interfaces and terminations \cite{botten2004,white2004,istrate2005,istrate2006}. The reflectance and the
penetration depth of the Bloch-mode mirrors of a microcavity follow from the same modes
\cite{sauvan2005,lalanne2008}, and in nanobeam cavities the mirror strength, the imaginary part of the Bloch
wavevector inside the gap, is the quantity by which the mirror segments are designed \cite{quan2011}. Between
two photonic crystal waveguides separated by $N$ rows the barrier enters the coupling through the $N$-th
power of an eigenvalue of its evanescent Bloch modes \cite{desterke2004}. For one point defect coupled to a waveguide, a Bloch-mode method with interface boundary
conditions reproduced the quality factor of a finite-difference time-domain (FDTD) computation to within a
few percent \cite{istrate2005jqe}. For an isolated cavity, the rates at which the spontaneous emission rate
and $Q$ grow with the number of surrounding layers correlate with the smallest decay constant of the
evanescent Bloch waves in the gap \cite{putra2011}. The coupling of a
cavity to a waveguide, on the other hand, is governed by the overlap of the evanescent cavity field with the
guided mode, and efficient couplers are designed for that overlap \cite{faraon2007}. For a cavity
side-coupled to a guide the two pictures point to different channels, the most slowly decaying one and the
one at the wavevector of the guided mode, and the question is which of them sets the factor gained per row.

Numerical methods agree less on the quality factor than on the resonance wavelength, and a benchmark of
five methods on photonic crystal membrane cavities found far larger disagreement on $Q$
\cite{delasson2018}. Perfectly matched layers lose their reflectionless property in periodic media, and
adiabatic absorbers replace them \cite{oskooi2008}. A photonic crystal waveguide of finite length acts as a
Fabry--P\'{e}rot resonator whose ends reflect the guided Bloch mode \cite{botten2003}. A mirror that
terminates a waveguide sets the $Q$ of a nanocavity coupled to it, and tuning the phase of the reflected
light changes that $Q$ dynamically \cite{tanaka2007}; partially reflecting elements in a guide make the
transmission line shape of a side-coupled cavity asymmetric \cite{fan2002}, as reflections at the facets of
an access waveguide do in experiments \cite{yang2007}. In a finite computational domain the same
reflections, and the leakage through a cladding of finite thickness, enter every computed quality factor.

Refractive-index sensing is a main application of such cavities: a defect mode shifts in frequency when the
index of the surrounding medium changes, by an amount set by the fraction of the mode energy in the analyte. Photonic crystal lasers probed femtoliter analyte volumes \cite{loncar2003}, and a year later
two-dimensional photonic crystal microcavities served as biochemical sensors \cite{chow2004};
silicon platforms for protein detection \cite{lee2007} and slotted cavities with integrated microfluidics
\cite{scullion2011} followed, and reviews cover photonic crystal resonances for sensing
\cite{pitruzzello2018}, point-of-care diagnostics \cite{inan2017} and nanophotonic biosensors in general
\cite{altug2022}. Line-defect-coupled L3 and H1-r nanocavities in perforated silicon
slabs give 63 and 155~nm per refractive index unit (RIU) \cite{dorfner2008}; slot cavities, which increase
the overlap of the mode with the analyte, reach more than 1500~nm/RIU with quality factors up to
$5\times 10^{4}$ \cite{difalco2009} and 510~nm/RIU at $Q \approx 2.6\times 10^{4}$ \cite{jagerska2010}. In
simulation a T-shaped slot waveguide gives 390--1040~nm/RIU \cite{turduev2017}; one-dimensional silicon nitride cavities
with asymmetric cladding sense in air and in liquids \cite{iadanza2022}, and waveguides side-coupled to two
microcavities give multiple Fano resonances \cite{harhouz2024}.

Sensor performance is quantified by the sensitivity $S = \dd\lamr/\dd\na$, the shift of the resonance
wavelength $\lamr$ per unit change of the analyte index $\na$, by the figure of merit
$\mathrm{FOM} = S/\Delta\lambda$, with $\Delta\lambda$ the linewidth, and by the detection limit (DL)
\cite{white2008}. The amplitude of the resonance matters as well as its width \cite{conteduca2022}, and the
waveguide geometry trades sensitivity against quality factor \cite{talebi2026}. In the near infrared,
absorption by an aqueous analyte limits $Q$; ultrahigh-$Q$ microcavities have used this absorption to detect
heavy water \cite{armani2006}. Deep-learning methods that design photonic structures
\cite{ma2021} or optimize the quality factor of photonic crystal nanocavities \cite{asano2018} learn from
computed quality factors and inherit any bias of the computation. A
sensor needs its operating wavelength and its linewidth set independently, and a side-coupled cavity offers
two parameters for this, the defect geometry and the barrier.

Breaking a mirror symmetry is a further means of control. In coupled-cavity waveguides in a square lattice
of rods of radius $0.20a$, auxiliary rods that break it raise the group index and the group-bandwidth product
of the guided mode \cite{oguz2024}; displacing the defect rod along the guide axis breaks
it in a side-coupled cavity. This study computes a cavity formed in the
$\Nsep$-th row of a square lattice of silicon rods, counted from a W1 waveguide, and varies the defect
radius, the displacement of the defect from its lattice site, the number of separating rows, the cladding
thickness, the mesh resolution and the termination of the guide. The factor gained per row is tested
against the evanescent channels of the bulk crystal, row by row and without a free parameter at two
operating points; the finite-domain biases of the computed quality factor are measured and removed; and the
sensitivity is measured over the index range 1.30--1.45 and compared with first-order perturbation theory.

% ================================================================= 2
\section{Methods}\label{sec:2}

The band structure and the guided-mode projection were computed by the plane-wave expansion (PWE) method
\cite{ho1990,johnson2001}, and the resonance frequency, the quality factor and the transmission spectrum by
the FDTD code Meep \cite{oskooi2010}. The open-source Side-coupled Photonic-crystal Resonator Analysis
Toolkit (SPRAT)\footnote{\spratrepo} holds the computational parameters and the analysis and plotting code, and the data are
openly available \cite{oguz2026data}.  SPRAT recomputes every reported number from the deposited files, and independent routes check
the results: the complex-band solver against the band edges of an established plane-wave code and by
convergence in its basis, the quality factor against the transmission spectrum, and the sensitivity against
perturbation theory.

\subsection{Physical model and geometry}\label{sec:2.1}

\begin{figure}[htbp]
\centering
\includegraphics[width=16cm]{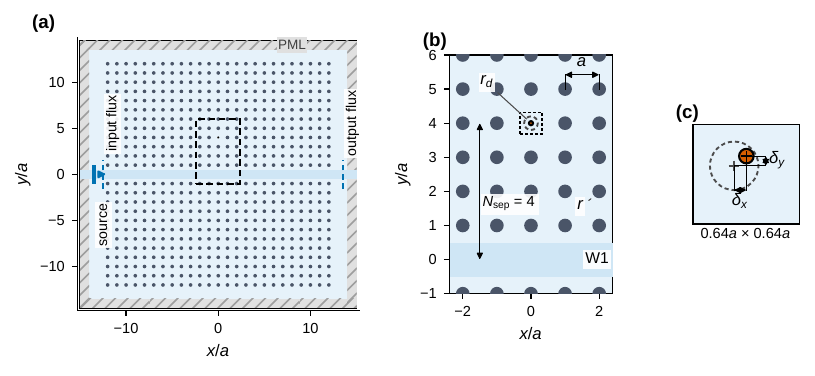}
\caption{Geometry of the two-dimensional photonic crystal microcavity with a side-coupled
point defect (reference design: $\rd = 0.060a$, $\Nsep = 4$, $\ncl = 12$). (a) The $30a \times 29a$
computational cell, the PML layers, the source and the flux planes; (b) the neighborhood of the cavity, with
the defect in the fourth row from the guide; (c) definition of the displacement components of the defect rod
(shown for illustration with $\dx = 0.10a$, $\dy = 0.08a$; in the reference design $\dx = \dy = 0$).}
\label{fig:1}
\end{figure}

The structure (figure~\ref{fig:1}) is a square lattice of lattice constant $a$ built from infinitely long
silicon rods of radius $r = 0.20a$; the refractive index of silicon is taken as $n_{\mathrm{Si}} = 3.45$
($\varepsilon = 11.9025$) and dispersion is neglected around 1550~nm. The space between the rods is filled
with the analyte, whose refractive index is denoted $\na$. Removing every rod along one lattice row forms a
W1 waveguide. Replacing one rod in the $\Nsep$-th row from the guide by a thinner rod of radius $\rd < r$
forms the side-coupled point defect \cite{villeneuve1996}; $\Nsep - 1$ complete rows then lie between
defect and guide, and $\Nsep$, which counts the row of the defect as well, is called the number of
separating rows. A defect formed by reducing a rod radius is an acceptor state: the mode is pulled from the
dielectric band up into the gap and concentrates its energy in the analyte around the defect. The
displacement of the defect rod from its lattice site along the guide axis is written $\dx$ and the
displacement perpendicular to the axis $\dy$. The model is two dimensional, with the rods taken to be
infinite perpendicular to the plane. Only the transverse-magnetic (TM) polarization, with the electric
field parallel to the rod axis, is treated, and out-of-plane losses fall outside the model.

\subsection{Band structure and the scaling of the lattice constant}\label{sec:2.2}

The TM and transverse-electric (TE) band structures of the bulk crystal were computed by plane-wave
expansion along the $\Gamma\text{--}X\text{--}M\text{--}\Gamma$ path. The evanescent channels inside the gap
were obtained from an in-house plane-wave expansion continued to complex wavevector
\cite{istrate2005,hsue2004}. At a fixed frequency and a fixed wavevector $k_x$ along the guide the TM
problem is quadratic in the transverse wavevector $k_y$, and linearizing it into a companion matrix turns
it into an ordinary eigenvalue problem. In a truncated basis part of its spectrum is spurious: those roots
have eigenvectors confined to the outermost reciprocal-lattice orders of the basis and move when its shape
changes. A root is kept as a Bloch channel when less than $10^{-3}$ of its eigenvector weight lies on the
two outermost orders; the solver returns each channel together with its replicas $k_y + 2\pi m/a$, with $m$
an integer, and the replica with $|\mathrm{Re}\,k_y| \le \pi/a$ is retained\supp{S6}. The replicas share
$\mathrm{Im}\,k_y$, so the decay constant does not depend on which one is kept. The decay constant of
the slowest channel changes by less than $2\times 10^{-4}$ between cutoffs of 7, 10 and 13
reciprocal-lattice orders in each direction and a circular basis of radius 16.

The lattice constant was calibrated to $\lambda_{\mathrm{t}} = 1550$~nm through the mid-gap frequency
$f_{\mathrm{mid}}$ by $a = f_{\mathrm{mid}} \cdot \lambda_{\mathrm{t}}$, and all computations use this
single lattice constant. Because the cavity mode lies below mid-gap, the results are rescaled once at the
end by $a \to a \cdot \lambda_{\mathrm{t}}/\lamr$, which brings the resonance to the target wavelength.
Under this transformation $Q$, FOM, DL and every normalized quantity are preserved exactly; only $\lamr$,
the linewidth and $S$ scale by the same factor.

\subsection{Computational domain and numerical parameters}\label{sec:2.3}

The computational cell contains 25 lattice periods along the guide and $2\ncl + 1$ periods across it, where
$\ncl$ is the number of cladding periods on each side of the guide; the number of cladding rows left above
the cavity, $\ncl - \Nsep$, is called the clearance. With $1.5a$ of padding along the guide, $1.0a$ across
it and a perfectly matched layer (PML) one period thick on all four sides, the cell, referred to as the
$25a$ cell, measures $30a \times (2\ncl + 5)a$, that is $30a \times 29a$ for $\ncl = 12$. Subpixel
smoothing reduces the discretization error at curved interfaces \cite{farjadpour2006}. The Courant number is
0.5, and at a resolution of 24 grid points per lattice period the cell with $\ncl = 12$ holds 501\,120 grid
points. When the defect is not displaced ($\dx = 0$) the structure has a mirror symmetry perpendicular to
the guide axis, and exploiting it halves the cell to be solved. The design with $\rd = 0.060a$,
$\Nsep = 4$, $\ncl = 12$, 24 points per period and $\dx = \dy = 0$ is the reference design. The number of rows needed above the cavity
grows with the quality factor to be measured, and the reference design has enough (section~\ref{sec:3.3}).

\subsection{Extraction of the resonance frequency and the quality factor}\label{sec:2.4}

The quality factor was obtained from the complex eigenfrequency returned by harmonic inversion (Harminv) of
the field sampled at a point inside the cavity after the source has been switched off
\cite{mandelshtam1997}; the width of the transmission dip serves as an independent check. The sampling time
$t$ of each computation was set from a preliminary estimate of $Q$ so that the sampling margin $\Qlim/Q$,
with
\begin{equation}
\Qlim = \pi f_{\mathrm{cen}}\, t
\label{eq:1}
\end{equation}
and $f_{\mathrm{cen}}$ the center frequency of the source, would reach a target of 1.50 to 4.50, set for each
parameter scan. $\Qlim$ is the quality factor of a mode whose amplitude falls by a factor $e$ during the
sampling time. Harmonic inversion is not limited by the Fourier resolution of the signal
\cite{mandelshtam1997} and resolves decay times longer than $t$, so the margin is a conservative sampling
criterion, and repeated computations at higher margin measure how conservative it is. The computation at
$\Nsep = 5$ with twelve cladding periods, limited by the longest sampling time used, reached a margin of
1.27; repetitions at margins of 2.54 and 4.78 reproduce its quality factor to 0.003~\%, and a repetition of
the reference design at a margin of 4.51 reproduces it to 0.001~\%\supp{S3}.

When harmonic inversion returns more than one mode, the mode of largest amplitude is taken as the
resonance. Modes close to a band edge extend over much of the finite
domain, so their quality factors depend on it. A mode therefore counts as valid when its frequency lies between 20~\% and 80~\% of the way
across the gap, the admission window, and its margin is at least 1. Of 858 computations, 17 fail the margin
condition and a further 100, all at defect radii of $0.110a$ and above, fall outside the window; the
remaining 741 form the data set, and none of the excluded ones belongs to the reference design. The
criterion tests the numerics alone. In a preliminary scan in a smaller domain ($\ncl = 6$, 20 points per
period), the mode of largest amplitude in 537 of the valid computations, those at $\Nsep = 1$ or
$\rd \ge 0.13a$, is a low-$Q$ mode of the finite waveguide whose wavelength does not follow the defect radius. These are
identified as waveguide modes and excluded from all fits\supp{S2}; the other valid computations hold the
cavity mode, and fits and statistics use only those. Excluded points that belong to a radius scan (the radius
$0.120a$ with eight cladding periods, and $0.110a$ in the scan of the reference design in steps of $0.005a$)
appear in the figures as open gray symbols. Acceptance criteria for the
verification computations were set before they were run; one was revised while they were under way, before
any cavity spectrum existed\supp{S7}.

\subsection{Coupling theory and line shape}\label{sec:2.5}

A single-mode cavity side-coupled to a through waveguide behaves as a notch filter in temporal coupled-mode
theory \cite{haus1984,fan2003}. In terms of the intrinsic decay time $\tau_{\mathrm{i}}$, the waveguide
coupling time $\tau_{\mathrm{w}}$ and the resonance frequency $\omega_{0}$ the transmission is
\begin{equation}
T(\omega) = \frac{(\omega-\omega_{0})^{2} + \tau_{\mathrm{i}}^{-2}}
                 {(\omega-\omega_{0})^{2} + (\tau_{\mathrm{i}}^{-1} + \tau_{\mathrm{w}}^{-1})^{2}}\,,
\label{eq:2}
\end{equation}
from which the minimum on-resonance transmission and the loaded quality factor follow as
\begin{equation}
T_{\min} = \left(\frac{Q_{\mathrm{L}}}{Q_{\mathrm{i}}}\right)^{2}, \qquad
\frac{1}{Q_{\mathrm{L}}} = \frac{1}{Q_{\mathrm{i}}} + \frac{1}{Q_{\mathrm{w}}}\,.
\label{eq:3}
\end{equation}
The measured pair ($Q_{\mathrm{L}}$, $T_{\min}$) determines $Q_{\mathrm{w}}$ uniquely through
$Q_{\mathrm{i}} = Q_{\mathrm{L}}/\sqrt{T_{\min}}$, provided the frequency grid resolves the dip. This
geometry differs qualitatively from the add-drop resonator \cite{manolatou1999}. There, critical coupling
imposes a trade-off, whereas in a side-coupled notch the dip deepens as $Q_{\mathrm{i}}$ grows and
$T_{\min} \to 0$ in the lossless limit. In that limit, adding separating rows therefore raises the loaded
quality factor without reducing the depth of the dip; at finite $Q_{\mathrm{i}}$ the same relation gives a
shallower dip as $Q_{\mathrm{w}}$ grows. Because the two-dimensional model has no out-of-plane radiation
channel, and twelve cladding periods make in-plane leakage small, $Q_{\mathrm{i}} \gg Q_{\mathrm{w}}$ and
$Q_{\mathrm{L}} \approx Q_{\mathrm{w}}$ are expected.

For a single-mode cavity side-coupled to a uniform guide, equation~(\ref{eq:2}) gives a symmetric
Lorentzian dip whatever the position of the defect \cite{fan2003}; partially reflecting elements in the
guide make it asymmetric \cite{fan2002}. The line shape was therefore fitted with the general Fano profile
\cite{limonov2017,fano1961}, $T = T_{\mathrm{bg}} + T_{0}\,(q + \xi)^{2}/(1 + \xi^{2})$ with
$\xi = 2(\lambda - \lamr)/\Delta\lambda$, in which a symmetric dip corresponds to $q = 0$. The transmission
is normalized by the output flux of a reference computation without the cavity. The flux read at the input
plane of the same computation collapses on resonance, because the notch reflects the whole incident power
back, so dividing by it narrows the dip artificially.

\subsection{Prediction of the sensitivity by perturbation theory}\label{sec:2.6}

For a small change in the refractive index of the analyte region, first-order perturbation theory gives
\begin{equation}
S = \frac{\dd\lamr}{\dd\na} \approx \frac{\lamr}{\na}\, \eta_{\mathrm{a}}\,,
\label{eq:4}
\end{equation}
where $\eta_{\mathrm{a}}$, the analyte energy fraction, is the fraction of the electric-field energy of the
mode that resides in the analyte. The fraction $\eta_{\mathrm{a}}$ was computed from the
discrete-Fourier-transform field at the resonance frequency and used as an independent check on the
sensitivity measured by FDTD.

\subsection{Finite-domain corrections}\label{sec:2.7}

Two properties of the finite domain enter every computed quality factor. The first is leakage through the
cladding. Energy leaves the cavity through the barrier toward the guide and also through the cladding on
the other side. The two channels add in the inverse quality factor, which gives the leakage law
\begin{equation*}
\frac{1}{Q_{N}} = \frac{e^{-\kappa N}}{A} + \frac{1}{Q_{\mathrm{top}}}\,,
\end{equation*}
where $Q_{N}$ is the quality factor with $N$ separating rows, $\kappa$ the barrier coefficient per row, $A$
a prefactor and $Q_{\mathrm{top}}$ the quality factor of the leakage alone, which depends on the clearance. Three computations at successive
row counts with the same number of rows above the cavity give three equations in the three
unknowns, with no fit and no assumed leakage. The same law removes the leakage wherever a corrected value
is quoted.

The second is reflection at the ends of the guide. The rod lattice, and with it the W1 guide, ends $12.5a$
from the cavity and $1.5a$ before the PML, and waves leaving the cavity return from these ends. A second
cell with 49 periods along the guide, the $49a$ cell, tests the dependence on the guide length. Continuing
the lattice and the guide for $8a$ into an adiabatic absorber \cite{oskooi2008} removes the abrupt end and suppresses the
reflection; the source and the output flux plane then lie inside the continued guide, half a
period from the absorber, and the domain is 41 periods long. The ratio of the quality factors with the two
terminations at matched geometry, after the cladding leakage is subtracted, is the cell factor
$G = Q_{\mathrm{PML}}/Q_{\mathrm{absorber}}$. With the leakage removed and the cell factor divided out, the
computed quality factor is the waveguide-coupling quality factor $Q_{\mathrm{w}}$ of
equation~(\ref{eq:3}) in an infinite guide, the reflectionless quality factor.

% ================================================================= 3
\section{Results}\label{sec:3}

\subsection{The photonic band gap and its dependence on the analyte index}\label{sec:3.1}

In water (nominal index $\na = 1.33$; about 1.32 at 1550~nm \cite{hale1973}, inside the index range
studied) a gap opens between the first and second TM bands over
$0.27526 \le a/\lambda \le 0.34585$, a relative width of 22.7~\%. No TE gap forms, as expected for a
rod-type geometry. The mid-gap frequency is 0.31055, so the lattice constant for 1550~nm is taken as
481.4~nm and the rod radius as 96.3~nm. With this lattice constant the gap corresponds to a 357~nm wide
wavelength window between 1391.8~nm and 1748.7~nm. The band edges agree within 0.1~\% with those of the
separate plane-wave expansion used for the complex band structure.

\begin{figure}[htbp]
\centering
\includegraphics[width=16cm]{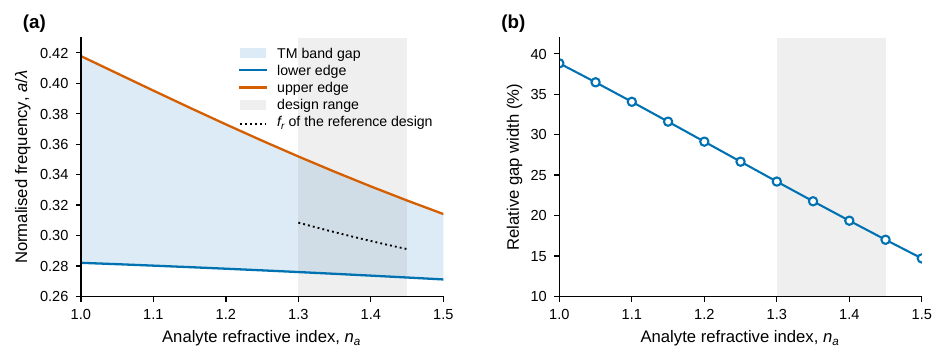}
\caption{(a) The lower and upper edges of the TM gap as a function of the analyte
refractive index; the dotted line is the resonance frequency $f_{\mathrm{r}}$ of the reference design over
the analyte index range and the shaded vertical band the design range. (b) The relative gap width as a
function of the analyte refractive index.}
\label{fig:2}
\end{figure}

Raising the analyte refractive index from 1.00 to 1.50 narrows the relative gap width from 38.8~\% to
14.7~\% (figure~\ref{fig:2}). The narrowing comes from the upper edge. The lower edge moves only 3.9~\%, from
0.2820 to 0.2711, while the upper edge falls 24.8~\%, from 0.4179 to 0.3141; the lower band concentrates its
energy inside the high-index rod and is little affected by the analyte, whereas the upper band carries a
substantial part of its energy in the analyte. Over the design range 1.30--1.45 the gap stays open, and in
the narrowest case the relative width is 17.0~\%.

\subsection{Continuous control of the resonance and discrete control of the quality factor}\label{sec:3.2}

\begin{figure}[htbp]
\centering
\includegraphics[width=16cm]{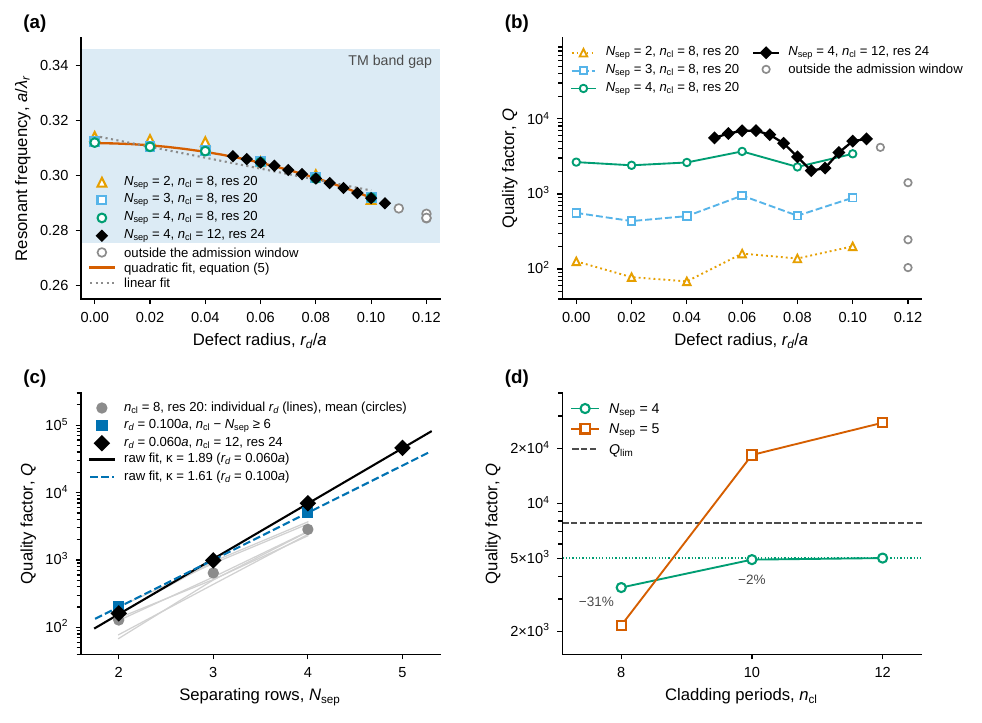}
\caption{(a) Resonance frequency against defect radius; the shaded area is the TM gap, the
solid line the quadratic fit of equation~(\ref{eq:5}) and the dotted line the linear fit. (b) Quality factor
against defect radius with eight cladding periods and for the reference design scanned in steps of $0.005a$.
(c) Quality factor against the number of separating rows (semi-logarithmic): gray lines and circles,
individual defect radii and their mean with eight cladding periods; squares, $\rd = 0.100a$ with at least six
cladding rows above the cavity; diamonds, $\rd = 0.060a$ with twelve cladding periods; solid line, log-linear
fit to the diamonds, and dashed line, the line through the squares, both on uncorrected quality factors
(``raw fit''). (d) Quality factor against the number of cladding periods ($\rd = 0.100a$, 32 points per
period); percentages are deviations from $\ncl = 12$ for $\Nsep = 4$, the dotted line is the $\Nsep = 4$
value at $\ncl = 12$, and the dashed line is $\Qlim$, above which the $\Nsep = 5$ values fail the sampling
criterion. In the legends ``res'' is the number of grid points per lattice period; in (a) and (b) open gray
circles mark modes outside the admission window.}
\label{fig:3}
\end{figure}

Changing the defect radius from $0.00a$ to $0.10a$ ($\Nsep = 4$, $\ncl = 8$, 20 points per period; at
$0.12a$ the mode lies outside the admission window) moves the resonance within the gap along a quadratic
law (figure~\ref{fig:3}(a)),
\begin{equation}
\frac{a}{\lamr} = 0.31165 + 0.0008\,\frac{\rd}{a} - 1.995\left(\frac{\rd}{a}\right)^{2},
\qquad R^{2} = 0.998\,,
\label{eq:5}
\end{equation}
with standard errors of 0.017 and 0.16 on the linear and quadratic coefficients. A linear fit leaves a
systematic curvature in its residuals ($R^{2} = 0.919$), and a cubic term brings no meaningful improvement
($R^{2} = 0.999$). The linear term is zero within its error, so $\Delta f \propto \rd^{2}$, as first-order
perturbation theory predicts: taking the completely removed rod ($\rd = 0$) as the reference, the
dielectric area put back at the lattice site is $\pi \rd^{2}$, and the frequency shift is proportional to
that area. Fitted with eight cladding periods at 20 points per period, the law predicts the twelve valid
points of a scan of the reference design in steps of $0.005a$ (twelve cladding periods, 24 points per
period) to within $2.2\times 10^{-4}$, about 1.1~nm at 1550~nm, so one law serves both domains. The quadratic
intercepts are 0.3141, 0.3119 and 0.3116 for $\Nsep = 2$, 3 and 4, a spread of 0.8~\%; the resonance
frequency depends only weakly on the number of separating rows.

The quality factor, by contrast, grows exponentially with the number of separating rows
(figure~\ref{fig:3}(c)):
\begin{equation}
Q \propto \exp(\kappa \Nsep)\,.
\label{eq:6}
\end{equation}
Measured without the cladding leakage and the end reflections (section~\ref{sec:2.7}), the barrier
coefficient $\kappa$ follows from two routes, which are independent apart from the small leakage correction
of the second. In the first, three quality factors computed with a PML termination and eight cladding rows
above the cavity give
\begin{equation}
e^{\kappa} = \frac{1/Q_{4} - 1/Q_{5}}{1/Q_{5} - 1/Q_{6}}\,,
\label{eq:6b}
\end{equation}
in which the leakage cancels, and so does the cell factor $G$ provided it is the same at the three row
counts; at $\rd = 0.060a$ this gives $\kappa = 1.9593$. In the second, the guide continues into the absorber, which suppresses the end reflections, and with the measured cladding leakage subtracted it gives
$\kappa = 1.9596$. Combining the absorber values at $\Nsep = 3$, 4 and 5 with the leakage-corrected and
cell-corrected value at $\Nsep = 6$ gives
\begin{equation*}
\kappa = 1.9595 \pm 0.002, \qquad 7.10 \text{ per row}\,,
\end{equation*}
with per-row steps of 1.9601, 1.9592 and 1.9593. The uncertainty is dominated by the discretization error,
propagated through the whole reduction, and the routes agree well within it. At $\rd = 0.100a$ the same
measurement, with an absorber and nine, eight and eight cladding rows above the cavity, gives
$\kappa = 1.739 \pm 0.007$. The barrier coefficient therefore depends on the operating point and has to be
quoted with the defect radius at which it was measured.

\begin{figure}[htbp]
\centering
\includegraphics[width=\textwidth]{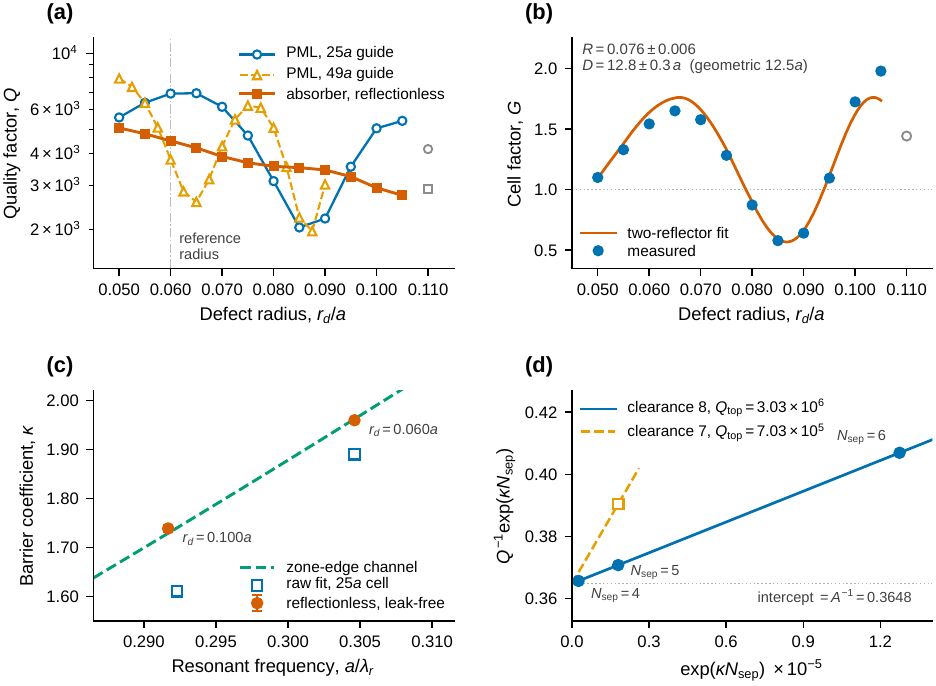}
\caption{(a) Quality factor against defect radius at $\Nsep = 4$ for three guide
terminations: the guides of the $25a$ and $49a$ cells, each closed by a perfectly matched layer, and the
absorbing termination, which suppresses the end reflections. (b) The cell factor
$G = Q_{\mathrm{PML}}/Q_{\mathrm{absorber}}$ at matched defect radius, with the cladding leakage removed, and
a two-reflector fit ($R$, reflectance; $D$, reflector distance). In (a) and (b) open gray symbols mark the
radius $0.110a$, outside the admission window. (c) Barrier coefficient against resonance frequency: the dashed
curve is the slowest evanescent Bloch channel of the complex band structure at $k_x = \beta$, at the zone
edge; filled circles, values without cladding leakage and end reflections (``reflectionless, leak-free'') at
$\rd = 0.060a$ and $0.100a$; open squares, the coefficients of the uncorrected quality factors of the $25a$
cell at the same radii (figure~\ref{fig:3}(c)). (d) $Q^{-1}\exp(\kappa \Nsep)$ against $\exp(\kappa \Nsep)$
for computations terminated by a perfectly matched layer with seven and eight cladding rows above the cavity
(``clearance''); the intercept is $A^{-1}$ and the slope $1/Q_{\mathrm{top}}$ of the leakage law
(section~\ref{sec:2.7}).}
\label{fig:4}
\end{figure}

Both values follow from the evanescent Bloch decay of the bulk crystal at the wavevector of the guided mode,
$k_x = \beta$, which is $0.5465\,\pi/a$ at $\rd = 0.060a$ and $0.4687\,\pi/a$ at $0.100a$\supp{S4}. The cavity couples to the guide only through the
component of its field that is phase-matched to the guided mode, so the barrier is crossed at that
wavevector. At both operating frequencies the slowest Bloch channel at $k_x = \beta$ lies at the zone edge,
$\mathrm{Re}\,k_y = \pi/a$, where the field alternates in sign from row to row. It gives 1.9620 and 1.7288
with no adjustable parameter (figure~\ref{fig:4}(c)), and the measured coefficients deviate from it by $-0.13$ and $+0.56$~\%. Both deviations are
comparable with the combined uncertainty of the measurement and of the prediction; the prediction inherits
the error of the frequency at which the channel is evaluated. The slower channel at
$k_x = 0$ would give a factor of 3.93 per row instead of 7.10, so the measurement distinguishes the
phase-matched channel from it. The next channel at $k_x = \beta$ decays by $e^{6.23}$ per row, and beyond
the first rows the zone-edge channel alone carries the tunneling. The channel fixes only the exponent; the
prefactor, and with it the absolute quality factor, still needs one computed point.

Coefficients taken from uncorrected computations come out low. With eight cladding periods the clearance
falls from six to four rows as $\Nsep$ grows from 2 to 4, and over the six valid radii the mean of
$\ln(Q_{4}/Q_{3})$ is $1.52 \pm 0.06$ (mean and standard error; table~\ref{tab:1}). Removing the cladding
leakage with the leakage law moves it to $1.90 \pm 0.04$ and reduces the scatter by a factor of 1.6. The law is then applied at five and four rows above the cavity, up to three rows beyond those at
which it was measured, and the individual corrected values are estimates. With twelve cladding periods and 24 points per period the quality factors at
$\Nsep = 2$, 3, 4 and 5 are 162, 987, 6927 and 46\,029 (figure~\ref{fig:3}(c)), and a log-linear fit gives
$\kappa = 1.890 \pm 0.019$. The $\Nsep = 2$ point is not yet asymptotic, the $\Nsep = 5$ point carries the
leakage through seven cladding rows, and every value carries the standing-wave factor of the finite guide.
At $\rd = 0.100a$ the uncorrected coefficient, the slope between $\Nsep = 2$ with eight cladding periods and
$\Nsep = 4$ with twelve, is 1.61.

The defect geometry sets the resonance wavelength and the barrier sets the order of magnitude of the
quality factor, so the two design degrees of freedom can largely be adjusted separately. The separation is
approximate. The barrier coefficient depends on the defect radius, the reflectionless quality factor falls
with radius (section~\ref{sec:3.8}), and in the $25a$ cell the quality factor also oscillates with radius,
by as much as a factor 3.4 (figure~\ref{fig:3}(b)).

\begin{table}[htbp]
\caption{Quality factor against the number of separating rows and the defect radius
($\ncl = 8$, 20 points per period, $\dx = \dy = 0$), with the logarithmic ratios of successive row counts. The
clearance above the cavity is five rows at $\Nsep = 3$ and four at $\Nsep = 4$; the last column is
$\ln(Q_{4}/Q_{3})$ with the cladding leakage removed by the leakage law (section~\ref{sec:2.7}). The radius
$0.120a$ (dagger) lies outside the admission window and is excluded from the means.}
\label{tab:1}
\centering\small
\setlength{\tabcolsep}{3.5pt}
\begin{tabular}{ccccccc}
\toprule
$\rd/a$ & $Q$ ($\Nsep = 2$) & $Q$ ($\Nsep = 3$) & $Q$ ($\Nsep = 4$) & $\ln(Q_{3}/Q_{2})$ & $\ln(Q_{4}/Q_{3})$ & \begin{tabular}[b]{@{}c@{}}$\ln(Q_{4}/Q_{3})$,\\ leakage removed\end{tabular} \\
\midrule
$0.000$ & $127$ & $559$ & $2650$ & $1.485$ & $1.557$ & $1.901$ \\
$0.020$ & $78$ & $435$ & $2404$ & $1.719$ & $1.709$ & $2.018$ \\
$0.040$ & $69$ & $506$ & $2622$ & $1.999$ & $1.645$ & $1.987$ \\
$0.060$ & $161$ & $949$ & $3679$ & $1.777$ & $1.355$ & $1.873$ \\
$0.080$ & $138$ & $513$ & $2287$ & $1.311$ & $1.495$ & $1.784$ \\
$0.100$ & $201$ & $889$ & $3419$ & $1.490$ & $1.347$ & $1.817$ \\
$0.120^{\dagger}$ & $105$ & $246$ & $1410$ & $0.853$ & $1.748$ & $1.917$ \\
\bottomrule
\end{tabular}
\end{table}

\subsection{Numerical convergence and finite-domain bias}\label{sec:3.3}

Between resolutions of 20 and 32 points per period the quality factor changes by at most 1.4~\% at four
defect radii (table~\ref{tab:2}). Tunneling through the intervening rows sets the quality factor and depends
little on the fine detail of the cavity, so it is stable even at low resolution. The resonance frequency
moves less still: by at most 0.023~\% between 20 and 32 points per period at the radii of table~\ref{tab:2},
and by 0.029~\% (0.46~nm at 1550~nm) between 24 and 32 at the reference radius ($\Nsep = 5$), although at 24
points per period the defect rod is only 1.44 grid points in radius. Subpixel smoothing preserves its area,
which sets the frequency at first order. The reference design uses 24 points per period, at
$(24/32)^{3} = 0.42$ of the cost at 32.

The cladding thickness limits the accuracy more (table~\ref{tab:3}, figure~\ref{fig:3}(d)). With four rows
left above the cavity the quality factor is 31~\% below the converged value, and with six rows 1.9~\% below;
eight rows, as in the reference design, is the last point of the scan. At 24 points per period
the difference between $\ncl = 10$ and $\ncl = 12$ is 2.9~\%, 3.9~\% and 1.5~\% for $\rd = 0.040a$, $0.060a$
and $0.080a$.

A fixed rule of the form $\ncl - \Nsep \ge 6$ fails at high quality factor, because the tolerable leakage
depends on the quality factor being measured. With eight cladding rows above the cavity and $\Nsep = 4$, 5
and 6 (figure~\ref{fig:4}(d)), the leakage law gives $Q_{\mathrm{top}} = 3.03\times 10^{6}$. The pair of
computations at $\Nsep = 5$ with seven and eight rows above the cavity gives $7.03\times 10^{5}$ for seven
rows, a decay of 1.461 per cladding row, between the decays of the evanescent channels at $k_x = 0$ and
$k_x = \pi/a$, as expected for leakage fed by a range of wavevectors along the guide\supp{S5}. The parasitic
channel carries 0.23~\% of $1/Q$ at $\Nsep = 4$, 1.60~\% at $\Nsep = 5$ and 10.34~\% at $\Nsep = 6$, all with
eight rows above the cavity, and 6.55~\% at $\Nsep = 5$ with seven. The number of rows required therefore follows
the target. Asking that $Q_{\mathrm{top}}$ exceed the barrier-limited quality factor by two orders of
magnitude gives seven rows above the cavity for the reference design, eight at the coupling optimum of
section~\ref{sec:4}, nine at $\Nsep = 5$ and ten at $\Nsep = 6$. The reference design, with twelve cladding
periods, meets it.

The second finite-domain effect is the guide itself. The ends of a finite photonic crystal guide reflect its
Bloch mode \cite{botten2003}, and for this guide a two-reflector fit gives $R = 0.076 \pm 0.006$ at a
reflector distance of $12.8 \pm 0.3a$ (figure~\ref{fig:4}(b)), against the geometric $12.5a$. Waves leaving
the cavity return and set up a standing wave at the cavity position, and the computed quality factor carries a factor set by the domain. At the reference radius the quality factors of the three terminations of
figure~\ref{fig:4}(a) differ by a factor of 1.83. The reflection arises at the end of the lattice, not in the
PML: doubling the thickness of the PML leaves the quality factor unchanged. The cell factor is 1.54 at the
reference design and flat to 0.02~\% between $\Nsep = 4$ and 5, and with the absorbing termination the
reflectionless quality factor of the reference design is 4502\supp{S4}. The absorbing termination
reproduces an infinite guide within its own uncertainty; a free overall scale in the fit of the ratio
against defect radius returns $0.978 \pm 0.024$.

In the $25a$ cell the quality factor also oscillates with defect radius (figure~\ref{fig:3}(b)). With eight cladding
periods the curves for two, three and four separating rows also vary with radius, but steps of $0.02a$
sample the oscillation too coarsely to compare their phases. The scan of the reference design in steps of $0.005a$ from $0.050a$ to $0.110a$ ($\na = 1.33$; the
last point lies outside the admission window) fixes the period. The quality factor is 5569 at
$\rd = 0.050a$, peaks at 6949 at $0.065a$, dips to 2037 at $0.085a$ and rises again to 5399 at $0.105a$, the
last valid radius. Peak and trough lie $0.020a$ apart, so the period is about $0.040a$, that is 19~nm. The
oscillation comes from the standing wave between the cavity and the guide ends (figure~\ref{fig:4}(a) and
(b)). Its period in frequency is inversely proportional to the guide length and to the group index of the
guided mode, and in the $49a$ cell the quality factor oscillates faster with radius (figure~\ref{fig:4}(a)).

\begin{table}[htbp]
\caption{Convergence of the quality factor with mesh resolution ($\Nsep = 4$, $\ncl = 8$).}
\label{tab:2}
\centering\small
\setlength{\tabcolsep}{4pt}
\begin{tabular}{cccc}
\toprule
$\rd/a$ & $Q$, 20~points/$a$ & $Q$, 32~points/$a$ & Relative difference (\%) \\
\midrule
$0.040$ & $2622$ & $2660$ & $+1.4$ \\
$0.060$ & $3679$ & $3731$ & $+1.4$ \\
$0.080$ & $2287$ & $2308$ & $+0.9$ \\
$0.100$ & $3419$ & $3469$ & $+1.4$ \\
\bottomrule
\end{tabular}
\end{table}

\begin{table}[htbp]
\caption{Convergence of the quality factor with the number of cladding periods
($\rd = 0.100a$, 32 points per period). The clearance is the number of lattice rows left above the cavity,
$\ncl - \Nsep$; the deviation is measured from the converged ($\ncl = 12$) value, and values above
$\Qlim = 7806$ fail the sampling criterion and are excluded from the data set.}
\label{tab:3}
\centering\small
\setlength{\tabcolsep}{4pt}
\begin{tabular}{cccccc}
\toprule
$\ncl$ & Clearance ($\Nsep = 4$) & $Q$ ($\Nsep = 4$) & Deviation (\%) & Clearance ($\Nsep = 5$) & $Q$ ($\Nsep = 5$) \\
\midrule
$8$ & $4$ & $3469$ & $-31.0$ & $3$ & $2155$ \\
$10$ & $6$ & $4926$ & $-1.9$ & $5$ & 18\,357 ($> \Qlim$) \\
$12$ & $8$ & $5024$ & $0.0$ & $7$ & 27\,527 ($> \Qlim$) \\
\bottomrule
\end{tabular}
\end{table}

\subsection{Displacement of the defect and radius of a barrier row}\label{sec:3.4}

\begin{figure}[htbp]
\centering
\includegraphics[width=16cm]{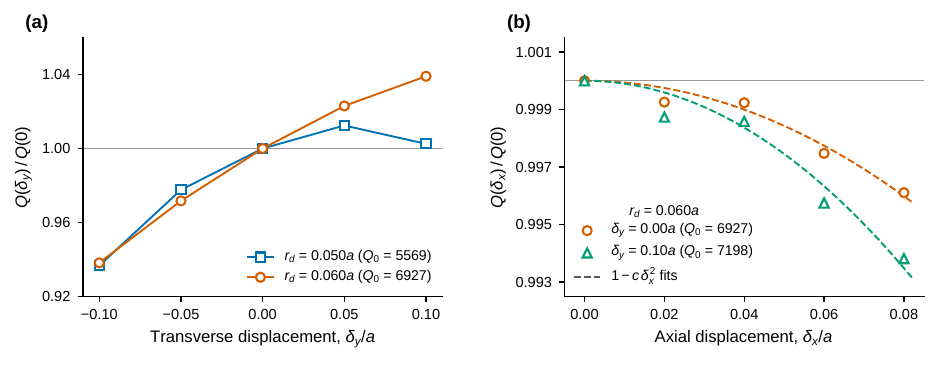}
\caption{Effect of displacing the defect rod from its lattice site on the quality factor
($\Nsep = 4$, $\ncl = 12$, 24 points per period), shown as the ratio to the quality factor $Q_{0}$ of the
undisplaced rod. (a) Displacement perpendicular to the guide axis, for two defect radii; (b) displacement
along the guide axis ($\rd = 0.060a$), for two values of $\dy$, on an expanded scale; the dashed lines are
fits of $1 - c\,\dx^{2}$.}
\label{fig:5}
\end{figure}

In the reference design the defect rod was displaced along both axes (figure~\ref{fig:5},
table~\ref{tab:4}). Displacement perpendicular to the guide axis raises the quality factor from 6498 at
$\dy = -0.10a$ to 7198 at $\dy = +0.10a$; a linear fit of $\ln Q$ against $\dy$ gives
$\kappa_{\delta y} = 0.512\, a^{-1}$ ($R^{2} = 0.978$), about one quarter (0.26) of the row-addition
coefficient. At $\rd = 0.050a$ the coefficient falls to $\kappa_{\delta y} = 0.342\, a^{-1}$
($R^{2} = 0.752$), the quality factor peaks at $\dy = +0.05a$ and turns back, and the best gain drops to a
factor 1.012. These values include the cell factor. The resonance shift that the displacement causes lowers
the cell factor by about 2~\% at $|\dy| = 0.10a$, equally for both signs, and leaves the linear coefficient
unchanged. The best displacement yields a factor 1.04 over the symmetric reference, far less than a row, so
the barrier acts in whole-row steps.

Changing the radius of the whole second barrier row, counted from the guide, through $0.18a$, $0.19a$,
$0.20a$, $0.21a$ and $0.22a$ changes the barrier itself and gives quality factors of 5664, 6376, 6927, 7326
and 7568, a smooth trim over about 15~\% of a row step in $\ln Q$, while the resonance wavelength moves by
2.1~nm. The cell factor, estimated from that shift and the slope of section~\ref{sec:3.8}, changes by about
4~\% over the range, against 34~\% for the quality factor.

Displacement along the guide axis lowers the quality factor at second order, by 0.39~\% at
$\dx = 0.08a$ with $\dy = 0$ and by 0.62~\% with $\dy = 0.10a$, following $Q(\dx)/Q(0) = 1 - c\,\dx^{2}$ with
$c = 0.63$ and $1.02\,a^{-2}$ to within 0.09~\% (figure~\ref{fig:5}(b)). The structure with the defect at
$-\dx$ is the mirror image of that at $+\dx$, so the quality factor and the resonance wavelength are even in
$\dx$. The decrease lies below the discretization error of an absolute quality factor, 1.4~\%, but a ratio at
fixed resolution cancels most of it, leaving a step of 0.03 to 0.06~\% between the symmetric computations at
$\dx = 0$, which used the mirror symmetry, and the displaced ones. Estimated from the resonance shift and the
same slope, the cell factor falls by 1.3~\% and 1.2~\% at $\dx = 0.08a$, more than the measured decrease, so
the intrinsic change is about one percent at most and its sign is not resolved.

The resonance wavelength is stationary at $\dx = \dy = 0$ and falls with the square of the displacement, by
1.1~nm at $|\dy| = 0.10a$ and 0.7~nm at $\dx = 0.08a$; a quadratic fit with a free vertex places the
stationary point at the lattice site, within what the 0.1~nm repeatability of the wavelength allows.

\begin{table}[htbp]
\caption{Effect of displacing the defect rod on the quality factor and the resonance
wavelength ($\Nsep = 4$, $\ncl = 12$, 24 points per period). The left block is for $\dx = 0$, the right block
for $\rd = 0.060a$.}
\label{tab:4}
\centering\small
\setlength{\tabcolsep}{4pt}
\begin{tabular}{ccccccc}
\toprule
$\dy/a$ & $Q$ ($\rd = 0.050a$) & $Q$ ($\rd = 0.060a$) & $\lamr$ (nm; $\rd = 0.060a$) & $\dx/a$ & $Q$ ($\dy = 0$) & $Q$ ($\dy = 0.10a$) \\
\midrule
$-0.10$ & $5216$ & $6498$ & $1579.2$ & $0.00$ & $6927$ & $7198$ \\
$-0.05$ & $5444$ & $6731$ & $1580.1$ & $0.02$ & $6922$ & $7189$ \\
$0.00$ & $5569$ & $6927$ & $1580.4$ & $0.04$ & $6921$ & $7187$ \\
$+0.05$ & $5638$ & $7086$ & $1580.1$ & $0.06$ & $6909$ & $7167$ \\
$+0.10$ & $5583$ & $7198$ & $1579.2$ & $0.08$ & $6900$ & $7153$ \\
\bottomrule
\end{tabular}
\end{table}

\subsection{Transmission spectrum and coupling regime}\label{sec:3.5}

\begin{figure}[htbp]
\centering
\includegraphics[width=16cm]{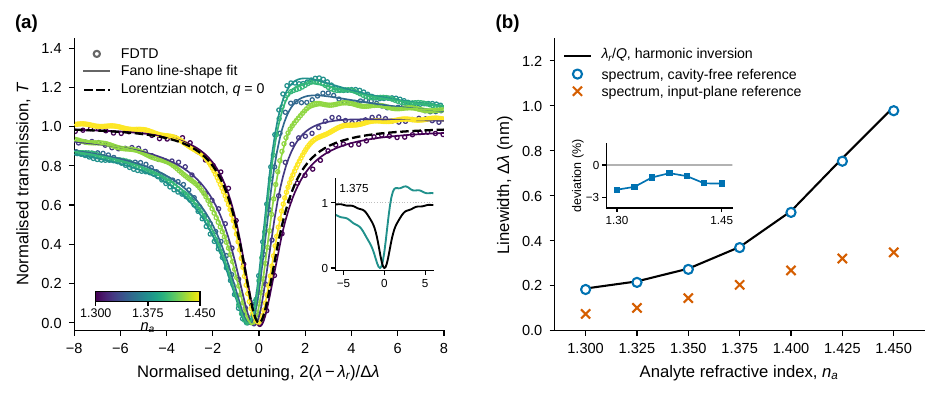}
\caption{(a) Normalized transmission at seven analyte indices against the detuning in units
of the half-width, $2(\lambda - \lamr)/\Delta\lambda$; the thin solid lines are Fano fits and the dashed line
the symmetric Lorentzian notch with $q = 0$; the inset shows $\na = 1.375$ in the $25a$ cell (color) and with
the absorber (black). (b) The linewidth $\Delta\lambda = \lamr/Q$ from harmonic inversion and from the spectra
normalized by the cavity-free reference computation; the crosses are the same spectra normalized to their
input plane. The inset is the percentage deviation of the spectral linewidth from the harmonic-inversion
value.}
\label{fig:6}
\end{figure}

The transmission spectrum was computed at seven analyte indices (1.300--1.450 in steps of 0.025) with 601
frequency points, with a separate reference computation without the cavity for each. Divided by the
reference, the transmission takes the value 0.992--1.005 more than fifteen linewidths from resonance, in the
five spectra that extend that far, within 1~\% of the value 1 expected for a lossless waveguide. Close to
resonance it exceeds unity by up to 25~\%, because the notch changes the round-trip phase between the
partially reflecting ends of the guide (figure~\ref{fig:6}(a)). Partially reflecting elements in the guide
make the line shape of a side-coupled cavity asymmetric \cite{fan2002}, as the facets of an access waveguide
do in experiments \cite{yang2007}; in the computation they are the truncated ends of the guide. Near their
minima the fitted dips are Lorentzian, with the asymmetry parameter $|q|$ between 0.03 and 0.50. With the
guide continued into the absorber, and the same frequency grid, source spectrum and criterion for ending the
time stepping, the
transmission near resonance stays below unity (at most 0.997) and $|q|$ falls to 0.024 or less, by a factor
of 8.95 or more at the five indices where it exceeded 0.2\supp{S4}. The overshoot, and the asymmetry where it
is large, therefore come from the finite guide ends.

Harmonic inversion gives the linewidth only through the definition $\Delta\lambda = \lamr/Q$. The spectrum
gives it independently, from the discrete-Fourier-transform flux and a line-shape fit. At the seven indices
the two values agree to better than 2.4~\% and the resonance wavelengths to within 0.005~nm
(figure~\ref{fig:6}(b)), which verifies the harmonic-inversion quality factor to that accuracy. The spectral
linewidth is narrower than the harmonic-inversion value at every index; the offset is common to the two
terminations and largely cancels in the cell factor. Normalized to the input plane of the same computation instead of to the reference computation, the
linewidth comes out narrower by a factor of 1.83 to 2.86 (the crosses of figure~\ref{fig:6}(b)).

The smallest sampled transmission, between $1.13\times 10^{-5}$ and $2.69\times 10^{-3}$ over the seven
spectra, is set by the frequency grid ($\Delta f = 6.0\times 10^{-6}$), and the samples next to it already lie
at $3\times 10^{-3}$ to $10^{-1}$. The inverse of equation~(\ref{eq:3}), $Q_{\mathrm{i}} =
Q_{\mathrm{L}}/\sqrt{T_{\min}}$, therefore returns numbers of order $10^{5}$--$10^{6}$ that follow the grid,
and at this resolution the depth of the dip cannot separate the intrinsic and waveguide-coupling components.
The measured cladding leakage does. In two dimensions it is the only intrinsic loss, so
$Q_{\mathrm{i}} = Q_{\mathrm{top}} = 3.03\times 10^{6}$ in the reference design, which puts the true minimum near $5\times 10^{-6}$. The loaded quality factor of the cell
is then its waveguide-coupling part to within the 0.23~\% that the leakage carries.

\subsection{Sensitivity, figure of merit and detection limit}\label{sec:3.6}

\begin{figure}[htbp]
\centering
\includegraphics[width=16cm]{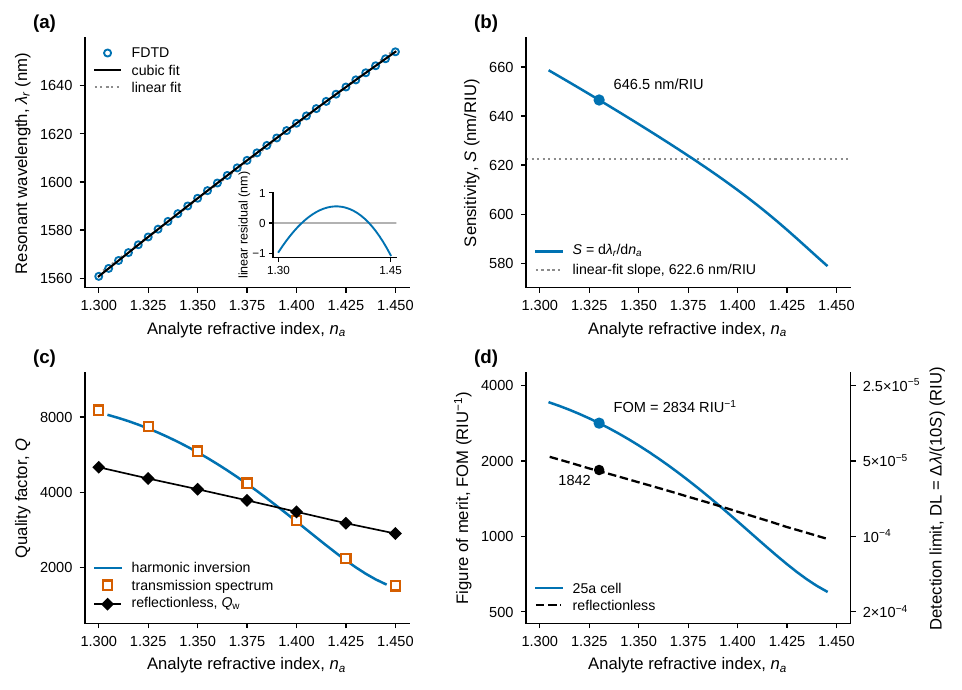}
\caption{Refractive-index sensing performance in a lossless medium. (a) Thirty-one-point
calibration curve with cubic and linear fits; the inset is the residual of the linear fit. (b) Local
sensitivity $S = \dd\lamr/\dd\na$ (central difference) compared with the slope of the linear fit. (c) Quality
factor against analyte index: the $25a$ cell by harmonic inversion (line) and from the transmission spectrum
(squares), and the reflectionless $Q_{\mathrm{w}}$ (diamonds). (d) Figure of merit and detection limit,
$25a$ cell (solid) and reflectionless (dashed); the dots mark the reference design.}
\label{fig:7}
\end{figure}

The resonance wavelength was computed at thirty-one points as the analyte index was varied from 1.300 to
1.450 in steps of 0.005 (figure~\ref{fig:7}(a)); every point satisfies the validity criteria (position inside
the gap 36.9--42.8~\%, margin 2.08--10.42). A single linear fit gives a slope of 622.6~nm/RIU and
$R^{2} = 0.9997$ but leaves a root-mean-square residual of 0.49~nm with a systematic curvature, so the
sensitivity is given as a local derivative:
\begin{equation*}
S(\na = 1.33) = 646.5\ \text{nm/RIU}, \qquad \frac{\dd S}{\dd\na} = -484\ \text{nm/RIU}^{2}\,.
\end{equation*}

The sensitivity decreases smoothly across the working range (table~\ref{tab:5}, figure~\ref{fig:7}(b)); these
values belong to the unscaled lattice constant ($a = 481.4$~nm, $\lamr = 1580.36$~nm). Under the scaling rule
of section~\ref{sec:2.2}, moving the resonance to 1550~nm sets $a = 472.2$~nm, $r = 94.4$~nm and
$\rd = 28.3$~nm, and shrinks the sensitivity and the linewidth by the same factor. The sensitivity of the
reference design becomes 634.1~nm/RIU and the linewidth of the $25a$ cell 0.224~nm. The figure of merit and
the detection limit are invariant under this scaling.

Over the same range the quality factor in the $25a$ cell falls from 8327 to 1662, a factor of 5.01
(figure~\ref{fig:7}(c)). Most of this fall comes from the cell factor, which drops from 1.65 at
$\na = 1.300$ to 0.61 at 1.450; the reflectionless value falls only from 5040 to 2740, a factor of 1.84, as
the index contrast and the gap narrow. At $\na = 1.330$, where both determinations of the reflectionless
quality factor exist, the value interpolated from the spectra lies 1.0~\% below the harmonic-inversion one.
Because the sensitivity decreases by 12~\% while the linewidth of the cell opens fivefold, against twofold
without the reflections, the figure of merit and detection limit of the cell depend strongly on the
operating point and the reflectionless ones far less (table~\ref{tab:5}, figure~\ref{fig:7}(d)). At
$\na = 1.33$, $\mathrm{FOM} = 2834$~RIU$^{-1}$ and $\mathrm{DL} = 3.53\times 10^{-5}$~RIU, with
$\mathrm{DL} = \delta\lambda_{\min}/S$ and the read-out resolution taken as
$\delta\lambda_{\min} = \Delta\lambda/10$. Under the convention $\delta\lambda_{\min} = \Delta\lambda/20$ the
same point gives $1.76\times 10^{-5}$~RIU.

Those values include the cell factor and assume a lossless medium. Without the reflections
(table~\ref{tab:5}) the figure of merit at the same operating point is 1842~RIU$^{-1}$ and the detection limit
is larger by a factor of 1.539; the absorption of water lowers the figure of merit further
(section~\ref{sec:4}).

\begin{table}[htbp]
\caption{Dependence of the performance quantities on the analyte index in a lossless medium
(unscaled lattice constant, $a = 481.4$~nm). $S$ is the local central difference and
$\mathrm{FOM} = SQ/\lamr$. $Q$ and FOM belong to the $25a$ cell; $Q_{\mathrm{w}}$ and
$\mathrm{FOM}_{\mathrm{w}}$ are reflectionless: harmonic inversion with the absorber at 1.330
(section~\ref{sec:3.3}), $Q$ over the cell factor of the spectra at 1.350 to 1.425, interpolated at 1.305
and 1.445.}
\label{tab:5}
\centering\small
\setlength{\tabcolsep}{5pt}
\begin{tabular}{ccccccc}
\toprule
$\na$ & $\lamr$ (nm) & $S$ (nm/RIU) & $Q$ & FOM (RIU$^{-1}$) & $Q_{\mathrm{w}}$ & $\mathrm{FOM}_{\mathrm{w}}$ (RIU$^{-1}$) \\
\midrule
$1.305$ & $1564.05$ & $658.4$ & $8137$ & $3425$ & $4938$ & $2079$ \\
$1.330$ & $1580.36$ & $646.5$ & $6927$ & $2834$ & $4502$ & $1842$ \\
$1.350$ & $1593.19$ & $636.7$ & $5778$ & $2309$ & $4114$ & $1644$ \\
$1.375$ & $1608.95$ & $623.8$ & $4323$ & $1676$ & $3714$ & $1440$ \\
$1.400$ & $1624.38$ & $609.8$ & $3057$ & $1148$ & $3343$ & $1255$ \\
$1.425$ & $1639.42$ & $593.6$ & $2138$ & $774$ & $3009$ & $1090$ \\
$1.445$ & $1651.15$ & $579.2$ & $1719$ & $603$ & $2792$ & $979$ \\
\bottomrule
\end{tabular}
\end{table}

\subsection{Analyte energy fraction and comparison with perturbation theory}\label{sec:3.7}

The discrete-Fourier-transform field of the resonant mode was collected in a separate computation for the
reference design in water. Near the rod boundaries subpixel smoothing mixes the two permittivities, and those
grid points, a tenth of the grid by number, hold 23~\% of the energy, because the acceptor mode concentrates
its energy at the rod edges; their treatment decides $\eta_{\mathrm{a}}$. Since $E_z$ is tangential to every
interface, the smoothed permittivity it sees is the arithmetic mean $\langle\varepsilon\rangle$ over the grid
cell, and the Hellmann--Feynman theorem applied to the discretized operator weights each boundary point by its
analyte fraction. Meep stores a different quantity, the harmonic mean of the eigenvalues of the smoothed
tensor. The tensor was therefore reconstructed from the geometry; it reproduces the stored map to $10^{-5}$,
and $\langle\varepsilon\rangle$ on the grid of $E_z$ follows from it\supp{S8}. The result is
$\eta_{\mathrm{a}} = 0.552$ and, through equation~(\ref{eq:4}), $S_{\mathrm{th}} = 656.3$~nm/RIU, 1.5~\% above
the measured 646.5~nm/RIU; the field map at $\na = 1.45$ gives 601.1~nm/RIU, 4.2~\% above the sensitivity
extrapolated to that index. Perturbation theory and the finite-difference computation therefore agree to
within a few percent. Counting the boundary points wholly as silicon or wholly as analyte brackets
$\eta_{\mathrm{a}}$ between 0.518 and 0.749, and a dielectric-threshold mask on the stored map gives
$\eta_{\mathrm{a}} = 0.618$ and overestimates $S$ by 13.6~\%.

\subsection{Fabrication tolerance}\label{sec:3.8}

The sensitivity of the resonance wavelength to the defect radius, taken from the scan of the reference design
in steps of $0.005a$ with a five-point derivative stencil, is $\partial \lamr/\partial \rd = 2.54$~nm/nm. This
agrees within 1.4~\% with the derivative of equation~(\ref{eq:5}) at the same point (2.57~nm/nm) and is
independent of the lattice constant by scale invariance. A fabrication error of $\pm 1$~nm in the defect
radius shifts the resonance by $\pm 2.54$~nm, about eleven linewidths in the finite cell and seven without
the reflections. The shift does not impair sensing, because it is small against the 357~nm gap and the
sensor measures the shift of the resonance; a tunable source or a single-point post-fabrication calibration
suffices, and only a read-out locked to a fixed-wavelength laser is excluded. For the quality factor a
linear coefficient becomes meaningful once the cell factor is removed. The oscillation with radius
(section~\ref{sec:3.3}) is a standing-wave feature of the finite guide: with an absorbing termination the
peak-to-peak variation of $\ln Q$ about a smooth trend is 10.2 times smaller, and the reflectionless quality
factor falls smoothly with defect radius. At the reference radius $\rd = 0.060a$ the local slope is
$\partial Q/\partial \rd = -120$~nm$^{-1}$, so a $\pm 1$~nm radius error changes the quality factor by
2.7~\%. In the $25a$ cell the slope is $+120$~nm$^{-1}$ (1.7~\% per nm), opposite in sign, because the
logarithmic derivative of the cell factor, $+21.6$ per $a$, outweighs the intrinsic $-12.9$ per $a$.

% ================================================================= 4
\section{Discussion}\label{sec:4}

The exponential growth of the quality factor with the rows around a point defect is long established
\cite{villeneuve1996}, and in channel-drop filters the same rows set the coupling
\cite{fan1998,fan1999,manolatou1999}. Evanescent Bloch modes have described the barrier between two coupled
waveguides \cite{desterke2004}, the quality factor of one waveguide-coupled defect \cite{istrate2005jqe}
and, for an isolated cavity in a hexagonal rod lattice, the growth of the emission rate and of the quality
factor with the number of layers, whose rates correlate with the smallest decay constant in the gap
\cite{putra2011}. In the side-coupled cavity the barrier
coefficient, measured free of the cladding leakage and of the end reflections, agrees with the decay
constant of the channel phase-matched to the guided mode, with no adjustable parameter, to
within 0.6~\% at two operating points. The channel at $k_x = 0$ would give a factor of 3.93 per row instead of 7.10. The
prescription for an isolated cavity, the most slowly decaying channel, decays no faster than that one, so it
does not carry over to a side-coupled cavity. The overlap of the cavity field with the guided mode, on which
scattering and coupled-mode analyses of cavity--waveguide coupling rest \cite{xu2000,waks2005} and for which
efficient couplers are designed \cite{faraon2007}, selects the channel, and the selection holds row by row.
The same evanescent Bloch modes underlie mirror design in nanobeam and other photonic crystal microcavities
\cite{sauvan2005,lalanne2008,quan2011}, and in photonic crystal slabs ultrahigh quality factors survive
barriers only a few rows thick \cite{kuramochi2008}.

Two verification steps apply to photonic crystal cavity computations in a finite domain: testing at what
cladding thickness the quality factor converges, and removing the reflections from the ends of the finite
waveguide. Numerical methods disagree far more on the quality factor than on the resonance wavelength
\cite{delasson2018}, and for this structure the barrier coefficient is 1.52 without converged cladding and
1.890 with twelve cladding periods but reflecting guide ends, against 1.9595. The reflection alone multiplies
the quality factor of the reference design by 1.54 and reverses the sign of $\partial Q/\partial \rd$. The leakage law, determined from a separate set of computations, moves the
uncorrected mean from 1.52 to 1.90, close to the reflectionless value, a test on computations from which it
was not derived. A mirror that ends a waveguide is a known means of setting the quality factor of a cavity
coupled to it \cite{tanaka2007}, and in a computation the abrupt end of the lattice is such a mirror.

The two-dimensional model assumes infinitely long rods and lossless dielectrics, and a real rod array adds two
losses, absorption in the aqueous analyte and radiation out of the plane. Linear interpolation in the absorption spectrum of water of Hale and Querry \cite{hale1973}, as tabulated by
Prahl \cite{prahl2018}, gives $\alpha \approx 10.8$~cm$^{-1}$ at 1550~nm\supp{S10}, an extinction coefficient
$k = \alpha\lambda/4\pi \approx 1.33\times 10^{-4}$. For a mode with a fraction $\eta_{\mathrm{a}}$ of its
electric field energy in an absorbing medium, the absorption-limited quality factor is
\begin{equation}
\Qabs \approx \frac{\na}{2k\,\eta_{\mathrm{a}}} = \frac{2\pi \na}{\eta_{\mathrm{a}}\,\alpha\,\lambda}\,.
\label{eq:7}
\end{equation}
The loss rate is the imaginary part of the same first-order shift that gives the sensitivity, so
equation~(\ref{eq:4}) fixes $\eta_{\mathrm{a}}$ from the measured $S$, and $\Qabs \approx \lamr/(2kS) \approx
9.2\times 10^{3}$ follows without a field map; because $k/\na \approx 10^{-4}$, the first-order treatment of
this loss is accurate. The measurements of Kou \textit{et al.} \cite{kou1993} give about 10~\% higher
absorption at 1550~nm and pull $\Qabs$ down to $8.3\times 10^{3}$. $\Qabs$ is of the same order as the
barrier-limited quality factor of the lossless model (4502). With $1/Q_{\mathrm{tot}} = 1/Q_{\mathrm{w}} +
1/\Qabs$, the total quality factor expected in water falls to 3020 and the figure of merit to
1235~RIU$^{-1}$, the detection limit grows by the same factor, and the sensitivity is unchanged. The three
figures of merit of the reference design, 2834 in the $25a$ cell, 1842 without the reflections and 1235 in
water, differ by more than a factor of two. A lossy analyte was not added to the FDTD computation. Meep smooths only the instantaneous permittivity at
interfaces, and with 23~\% of the energy at the boundary points such a computation would test the grid
rather than the first-order formula.

Raising the barrier to $\Nsep = 5$ multiplies $Q_{\mathrm{w}}$ by 7.10, and the total quality factor is then
limited by $\Qabs$. It reaches 7127 against 3020 at $\Nsep = 4$, while the notch becomes shallow, its minimum
transmission rising from 0.108 to 0.603. The amplitude of a resonance matters for detection as well as its
width \cite{conteduca2022}, and for a notch the product $Q_{\mathrm{tot}}(1 - T_{\min})$ is adopted here as a measure of a line that is
both narrow and deep. With $Q_{\mathrm{i}} = \Qabs$ in equation~(\ref{eq:3}) it peaks at $Q_{\mathrm{w}} = \tfrac{1}{2}(1+\sqrt{3})\,\Qabs \approx 1.25\times
10^{4}$, and by this measure the two designs are nearly equivalent, at 0.763 and 0.800 of the maximum. The
figure of merit, which depends only on the linewidth, prefers $\Nsep = 5$ by a factor of 2.4, 2916 against
1235~RIU$^{-1}$, whereas a read-out limited by the depth of the notch prefers $\Nsep = 4$. The optimum lies
between the two row counts and outside the range that the radius of a barrier row covers, so the choice
between them rests on the read-out.

A finite-height realization adds a vertical radiation channel, and the three losses combine as
\begin{equation}
\frac{1}{Q_{\mathrm{tot}}} = \frac{1}{Q_{\mathrm{w}}} + \frac{1}{Q_{\perp}}
+ \frac{1}{\Qabs}\,.
\label{eq:qperp}
\end{equation}
Relative to the 3020 set by the barrier and the medium, a vertical channel lowers $Q_{\mathrm{tot}}$ by
2.9~\% for $Q_{\perp} = 10^{5}$, by 10~\% for $2.7\times 10^{4}$ and by 23~\% for $10^{4}$. For the present
guide the vertical channel is not a small correction. At both operating points the propagation constant of
the guided mode is smaller than the free-space wavenumber, so in a slab of finite height the guided mode, and
with it the phase-matched component of the cavity field, lies inside the light cone of any cladding and radiates. Only modes outside the light cone are
truly guided by a line defect in a photonic crystal slab \cite{johnson2000}. In a slab of rods the band gap lies in transverse-magnetic-like modes, whose effective
dielectric constant is far lower than that of the transverse-electric-like modes of a perforated slab, so the
gap of a rod slab is widest at a height near $2.3a$ against $0.6a$ for a hole slab, both for a dielectric
constant of 12 in air \cite{johnson1999}. The two-dimensional results therefore describe the limit of long
rods. A slab design needs a guide whose mode lies below the light line, and its per-row factor has to be
computed from the complex band structure of the slab at that propagation constant.

A rod-type geometry is hard to fabricate as a suspended structure, but pillar cavities of the
transverse-magnetic kind exist: SiO$_2$/Si nanopillars 210~nm across on silicon-on-insulator reach
$Q \approx 4\times 10^{3}$ and have been integrated beside an on-chip waveguide \cite{poblet2025}, against a
defect rod 57~nm across at 1550~nm. The equivalent hole-type silicon slab is easier to make, but its mode
would likely hold a smaller analyte energy fraction and so lose sensitivity. The barrier-limited quality
factor, 4502, lies above those of perforated-slab nanocavity experiments \cite{dorfner2008} and below those of
slotted cavities \cite{difalco2009,jagerska2010}. The sensitivity of 634~nm/RIU lies above the perforated-slab
values and the 510~nm/RIU of an air-slot cavity and below the more than 1500~nm/RIU of a slotted
heterostructure cavity. Experimental and simulated values are not directly comparable\supp{S11}.

% ================================================================= 5
\section{Conclusions}\label{sec:5}

The resonance wavelength of a point-defect microcavity side-coupled to a W1 waveguide in a two-dimensional
photonic crystal of silicon rods in water is set continuously by the defect radius, and the quality factor,
in steps of a factor 7.10 at the reference radius, by the number of lattice rows between cavity and
waveguide. The barrier coefficient, the exponent per row, agrees with the decay
constant of the evanescent Bloch channel of the bulk crystal that is phase-matched to the guided mode,
which lies at the zone edge: 1.9620 and 1.7288 predicted against
1.9595 and 1.739 measured. The channel at zero wavevector along the guide would give a factor of 3.93 per
row instead of 7.10. Displacing the defect rod by up to a tenth of a lattice period changes the quality
factor by less than 7~\%, while the radius of a barrier row trims it within a step. Scaled to 1550~nm the
reference design gives a barrier-limited quality factor of 4502, a sensitivity of 634~nm/RIU and a figure of
merit of 1842~RIU$^{-1}$ in a lossless medium, or 1235~RIU$^{-1}$ with the absorption of water, which also limits the
return on adding rows. Perturbation theory evaluated on the discretized structure reproduces the sensitivity
to 1.5~\%. In a finite computational domain the barrier coefficient and the quality factor can be trusted only after convergence in cladding thickness and removal of the reflections from the guide ends;
without them this structure gives 1.52 or 1.890 in place of 1.9595.

The guided mode lies inside the light cone at both operating points, so a slab of finite height needs a
different guide. A guide whose mode lies below the light line, a three-dimensional computation with a finite
rod height and a substrate, and the cost in sensitivity of moving to a hole-type silicon slab are the next
steps toward fabrication. The data set can
also serve wider scans of the defect geometry and the training of surrogate models.

% ================================================================= BACK MATTER
\section*{Acknowledgments}

The author gratefully acknowledges the support of the Material Physics Simulation Laboratory at Pamukkale
University. This work was supported by the Istanbul Okan University Scientific Research Projects Coordination
Unit (IOU BAP) under Project No.\ OBAP2026010006, and by the Pamukkale University Scientific Research Projects
Coordination Unit (PAU BAP) under Project No.\ 2025ALDEP037. Computing resources were provided by the National
Center for High Performance Computing of Turkey (UHeM) under grant number 5027772026. 

\section*{Author Declarations}

\subsection*{Conflict of Interest}

The author has no conflicts to disclose.

\subsection*{Author Contributions}

\textbf{Hasan Oguz:} Conceptualization (lead); Data curation (lead); Formal analysis (lead); Funding
acquisition (lead); Investigation (lead); Methodology (lead); Software (lead); Validation (lead);
Visualization (lead); Writing -- original draft (lead); Writing -- review \& editing (lead).

\section*{Data Availability}

% Ragged right: the two URLs stay whole and no line of the paragraph is stretched.
{\raggedright
The data that support the findings of this study are openly available in Zenodo at \zenodolink, reference
\cite{oguz2026data}. They comprise the harmonic-inversion results with their geometry parameters and
validity flags, the transmission spectra with their reference computations, the stored field maps, the
band-structure computations, the acceptance criteria of the verification computations with their outcomes,
the analysis outputs, the scripts that produced the data and the supplementary document. The same Zenodo
entry holds SPRAT version 1.2.1, the software with which the data were converted and analyzed; SPRAT is
maintained at \spratrepo.\par}

% ================================================================= REFERENCES

\end{document}